\documentstyle[prd,aps]{revtex}
\begin{document}

\draft
\preprint{
UTEXAS-HEP-97-19
UMD-PP-98-07
SMU-HEP 97-12
{}}
\title{Supernova Constraints on a Superlight Gravitino}
\author{Duane A. Dicus$^{1}$, Rabindra N. Mohapatra$^{2}$ and
Vigdor L. Teplitz$^{3}$}
\address{$^{(1)}${\it Department of Physics, University of Texas
at Austin, Austin, TX-78712.}}
\address{$^{(2)}${ Department of Physics, University of Maryland, 
College Park, MD-20742, USA.}}
\address{$^{(3)}${ Department of Physics, Southern Methodist University,
Dallas, TX-75275}}
\date{August, 1997}
\maketitle
\begin{abstract}
In supergravity models with low supersymmetry breaking scale
the gravitinos can be superlight, with mass
in the $10^{-6}$ eV to few keV range. In such a case,
gravitino emission provides a new cooling mechanism for protoneutron
stars and therefore can provide constraints on the mass of a
superlight gravitino. This happens because the  coupling to matter 
of superlight gravitinos is dominated by its goldstino
component, whose coupling to matter is inversely
proportional to the scale of supersymmetry breaking and inceases as
the gravitino mass decreases. Present observations therefore
provide lower limits on the gravitino mass. Using the recently revised
goldstino couplings, we find that the two dominant processes
in supernova cooling are $e^+e^-\rightarrow \tilde{G}\tilde{G}$ and
$\gamma+e^-\rightarrow e^-\tilde{G}\tilde{G}$. They lead to a lower
limit on the supersymmetry breaking scale $\Lambda_S$ from 160 to
500 GeV for core temperatures 30 to 60 MeV and electron chemical
potentials 200 to 300 MeV. The corresponding 
lower limits on the gravitino mass are $.6-6\times
10^{-6}$ eV. 
             
\end{abstract}
\pacs{\hskip 2 cm UTEXAS-HEP-97-19 \hskip 1 cm UMD-PP-98-07 \hskip 1 cm 
SMU-HEP-97-12}

\vskip2pc

\section{Introduction}
In supergravity models with a low supersymmetry breaking scale,
the gravitino ($\tilde{G}$) mass is in 
the superlight range of $10^{-6}$ eV to
few keV's since it is given by the formula 
$m_{\tilde{G}}\sim\frac{ \Lambda^2_{\rm S}}{ 2\sqrt{3}M_{P\ell}}$
(where $\Lambda_{\rm S}$ is the scale of supersymmetry breaking and
$M_{P\ell}$ is the Planck mass that characterises the gravitational
interactions.)
Any information on the gravitino mass therefore translates into knowledge
of one of the most fundamental parameters of particle physics, the scale
at which supersymmetry breaks.

There is an advantages in studying the gravitino's properties
when its mass is in the superlight range because  it can be
easily emitted in astrophysical processes such as
supernova cooling\cite{grifols}, neutron star cooling etc. This adds new
cooling mechanisms to the already known ones for supernovae, i.e. the usual
neutrino emission, which the observation of neutrinos from  SN1987A
seems to have confirmed. Any additional process can therefore afford a
maximum luminosity of roughly $10^{52}$ ergs/sec. This will then lead to
constraints on the parameters that describe the coupling of the gravitinos
to matter in the supernovae. Furthermore since for light gravitinos
one has $\tilde{G}_{\mu}\simeq i\sqrt{\frac{2}{3}}\frac{1}{m_{\tilde{G}}}
\partial_{\mu}\chi$,
the superlight gravitino coupling is dominated by the coupling of the
goldstino to matter, which is inversely proportional to the supersymmetry
breaking scale-squared $F\equiv \Lambda^2_{\rm S}$. For values of $\Lambda_{\rm
S}$ in the 100 GeV to TeV range, the goldstino coupling strengths to matter
are of the same order as the ordinary weak interactions. The gravitino emission
in astrophysical settings such as the supernovae can therefore be competitive
with the neutrino emission process. Observed supernova neutrino luminosity 
by the IMB and Kamiokande
groups \cite{bionta} and its understanding in terms of the standard model
of the supernova \cite{burrows} therefore 
allows us to set lower limits on 
$\Lambda_{\rm S}$ and hence on $m_{\tilde{G}}$.

The supernova and other astrophysical constraints on the gravitino mass 
were first studied in two 
recent papers\cite{grifols,nowa,riotto}. The first paper considered the class
of models where the superlight gravitino is the only superlight particle
in the model with  its superpartners having masses in the GeV range
whereas  the last two papers considered the smaller subclass of models
\cite{ellis} where the gravitino is
also accompanied by superlight scalar and pseudoscalar particles. In this paper
we will focus on the first class of models.

In Ref.\cite{grifols}, gravitino couplings to matter suggested  
in Ref.\cite{gherghetta} were used. These
couplings have recently been criticized in two papers\cite{cern,luty}
and a new set of matter couplings have been proposed for the
class of supersymmetry models where the scalar and pseudoscalar
partners of the gravitino are heavier (e.g. in the multi GeV range)
 than the gravitino. Since
the temperature dependence of gravitino emission rates are very different
if one uses the new set of Feynman rules for gravitino matter couplings, it
is necessary to revisit these bounds again. It
is the goal of this paper to use the revised Feynman 
rules for matter gravitino coupling to calculate  gravitino emission
rates in supernovae,  obtain the bounds on the supersymmetry breaking 
scale, and from them the lower limit on the gravitino mass\footnote{As this
paper was being prepared for publication, a paper by Grifols et 
al\cite{toldra} appeared which used the revised Feynman rules to obtain
a lower limit on the gravitino mass. While our final results are very
similar, we find two processes dominating  gravitino 
emission process whereas  Ref.\cite{toldra} considers only one of them.}
Our result is that the basic processes that dominate the energy loss via
gravitino emission are $e^+e^-\rightarrow \tilde{G}\tilde{G}$
(called annihilation process below) and
$\gamma +e^-\rightarrow e^-\tilde{G}\tilde{G}$,
(called Compton process below) whereas the process
found to dominate in Ref.\cite{grifols} i.e. $\gamma\gamma\rightarrow
\tilde{G}\tilde{G}$ is found to make a negligible contribution. The physically
interesting lower limit on the $m_{\tilde{G}}$ remains $\sim 3\times 10^{-6}
\times\left(\frac{M}{100~GeV}\right)^{1/2}$ eV (where $M$ is a model
dependent parameter, which can lie anywhere from 50 to 250 GeV).
This bound is qualitatively somewhat better than 
the one found in Ref.\cite{grifols}.

This paper is organized as follows: in section II, we note the various possible
processes that can contribute to energy loss from supernovae via $\tilde{G}$
emission and, using simple dimensional analysis, obtain the 
temperature dependence of
the emissivity for the different processes and give qualitative
arguments to isolate the processes that dominate the emissivity; 
we calculate the emissivity ($Q$)
for the Compton process in sec. III 
and for the annihilation processes in sec. IV and derive a lower bound
on $\Lambda_{\rm S}$ and $m_{\tilde{G}}$ from supernova; 
in section V we discuss the energy loss from neutron stars.
                                         
\bigskip
\section{Dominant mechanisms for energy loss via gravitino emission}

Since gravitinos are superpartners of gravitons, which have universal coupling
to matter, we expect them to couple to all matter in pairs. Specifically,
 in the
case when the gravitino is  superlight, the dominant couplings arise from
the coupling of its longitudinal mode\cite{fayet}. Goldstinos, like Goldstone
bosons, must be derivatively coupled. Combining this property with the
constraints of supersymmetry, Luty and Ponton\cite{luty} have derived
the form of the effective coupling between matter and goldstinos. The 
couplings of interest to us are given by\cite{luty}
\begin{eqnarray}
{\cal L}_{\gamma\chi\chi}=\frac{eM^2}{\Lambda^4_{\rm S}}
\partial^{\mu}\bar{\chi}\gamma^{\nu}\chi F_{\mu\nu}
\end{eqnarray}
\begin{eqnarray}
{\cal L}_{\chi\psi\tilde{\psi}}=\frac{2}{\Lambda^2_{\rm S}}
\partial^{\mu}\bar{\chi}\psi(D_{\mu}\tilde{\psi})
\end{eqnarray}
\begin{eqnarray}
{\cal L}=\frac{-i\sqrt{2}}{\Lambda^2_{\rm S}}
\partial^{\mu}\bar{\chi}\gamma^{\nu}\lambda_AF_{\mu\nu A}
\end{eqnarray}
where we have denoted the goldstino by $\chi$ and the generic gaugino
by $\lambda_A$ for the gauge field $A$. In the $\gamma\chi\chi$
coupling the mass parameter $M$ is highly model dependent
and sensitive to the details of the supersymmetry breaking sector
as well as the nature unification group\footnote{
For instance in grand unified theories with simple groups, at the
GUT scale $M^2=0$. Therefore at low energies, this will give additional
suppression factors}. In Ref.\cite{luty}
a value of 43 GeV is suggested assuming the typical slepton mass to be
200 GeV and the squark mass to be 400 GeV. But it is perfectly possible
even in the context of gauge mediated models to have very different parameters
e.g. slepton masses of 200 GeV and squark masses to be a TeV\cite{chacko}.   
We have therefore kept this as a free parameter and express all our results
in terms of this number.

We are now ready to discuss the detailed processes that can contribute to
the energy loss in in a supernova. There are the following classes of processes:
(i) ones that involve both nonrelativistic particles (e.g. neutrons and
protons) and relativistic particles such as $e^{\pm}$, $\gamma$ and $\chi$;
 (ii) that involve only relativistic particles; and finally, (iii) 
the ones that involve the decay 
of the plasmon i.e. $\gamma^*\rightarrow \tilde{G}\tilde{G}$. 

A typical
process of type (i) is the analog of the modified URCA process for neutrino
emission with neutrinos replaced by the gravitinos : $NN\rightarrow
NN\chi\chi$. Even though the gravitino emission graph involves the
exchange of a virtual photon, the derivative couplings present in
the gravitino vertex imply that the photon-gravitino-gravitino coupling
in momentum space goes like $k_1\cdot k_2= (k^2-2m^2_{\tilde{G}})/2$ where
$k=k_1+k_2$ and $k_i$ are the gravitino momenta. For superlight gravitinos,
the effective virtual photon induced $e^+e^-\rightarrow \tilde{G}\tilde{G}$
amplitude is like a pointlike four Fermi coupling with $G_F/\sqrt{2}$
replaced by $e^2M^2/8\Lambda_{\rm S}$. One can therefore use this feature
to calculate the gravitino pair emissivity from the known
the URCA processes\cite{friman}.
One can conclude that the emissivity $Q$, defined as the amount of
energy emitted from the supernova per unit volume per unit time 
($E^5$ in natural units) is given by
\begin{eqnarray}
Q_{NN}\simeq (164\pi^3/4725)\alpha^2\alpha^2_{\pi}(M/\Lambda^2_{\rm S})^4
p_FT^8
\end{eqnarray} 

Let us now consider several examples of the class (ii) processes.
They are (a) $e^+e^-\rightarrow \tilde{G}\tilde{G}$; (b) $\gamma +e^-
\rightarrow e^-+\tilde{G}+\tilde{G}$; (c) $\gamma +\gamma\rightarrow
\tilde{G}\tilde{G}$. Let us first focus on the process (c) which is
the weakest of the three. Since this process involves  exchange
of a photino, its coupling strength is given by $\sim 
e^2/(\Lambda^4_{\rm S}M_{\lambda_{\gamma}})$. Therefore from simple
dimensional arguments we conclude that the emissivity for this process
will go like $\sim (\alpha^2/\Lambda^8_{\rm S}M^2_{\lambda_{\gamma}})T^{15}$.
Let us now apply similar dimensional arguments to the process (b). In this case
the Feynman diagrams that contribute are given in Fig. 1. Note that
all exchange particles are also light. Therefore the only dimensional
parameters are provided by the photon-gravitino coupling strength
$M^2/\Lambda^4_{\rm S}$. Using dimensional analysis,
one gets the following temperature dependence for $Q$: $\sim\alpha^3 
\left(\frac{M}{\Lambda^2_{\rm S}}\right)^4 T^9$. Here, we have ignored the
effects of Pauli exclusion principle in this estimate; this will be
addressed below. But assuming that this effect
does not lead to a drastic change in temperature dependence, we see that
it clearly dominates over the process (c) since
supernova core temperatures are of order 30 to 60 MeV. 

Process (a) is more subtle since the electrons are degenerate in the 
supernova core. As a result of this the positron density is suppressed
compared to that of the electron by
by $(T/\mu)^3e^{-\mu/T}$, 
where $\mu$ is the chemical potential for the electrons.
Therefore even though the naive dimensional arguments imply that for this
process, the emissivity goes like $\alpha^2\left(\frac{M}{\Lambda^2_{\rm S}}
\right)^4 T^9$ and is therefore more dominant than the process (b), we will
see that in actual practice they are comparable. 

The plasmon decay has been considered in \cite{toldra} and is found to
be negligible. We therefore do not discuss it here.

\bigskip
\section{Calculation of the emissivity via the Compton process}
The generic formula for the emissivity $Q$ for a process with
initial particles and momenta denoted by $i_a$ and $p_{i_a}$ and
final state particles and momenta denoted by $f_a$ and $p_{f_a}$ is:
\begin{eqnarray}
Q=\int \Pi_{i,f}\frac{d^3p_{i_a}}{2E_{i_a}}\frac{d^3p_{f_a}}{2E_{f_a}}
({2\pi})^{4-3n_i-3n_f}\\ \nonumber
 \delta^4(\Sigma p_{i_a}-\Sigma p_{f_a}) 
\Pi_{i_a}f(p_{i_a})\Pi_{f_a}(1-f(p_{f_a})) |M_{i,f}|^2 (E_{f1}+E_{f2})
\end{eqnarray}
In the equation above, $M_{if}$ denotes the matrix element for the process
responsible for  gravitino emission; $E_{f1,f2}$ are the energies of the 
gravitinos; $f(p)$ are the thermal distributions for
 the various particles involved in the process. For example the thermal 
distribution of the electrons and  positrons are given by:
\begin{eqnarray}
f_{e^\mp}= 1/ (e^{\frac{E\pm \mu_e}{T}}+1)
\end{eqnarray}
where $\mu_e$ is the chemical potential for the electrons. Similarly for
the photons we have $f_{\gamma}(p)=1/(e^{\frac{E}{T}}-1)$. 
For the gravitinos, we
choose the $f$'s to be zero since they do not get a chance to thermalize
in the supernova.

Let us apply this formula to the Compton process which is given by the two
Feynman diagrams in Fig.1. Let us denote the initial and final momenta as 
follows:
\begin{eqnarray}
\gamma(q)+e(p_1)\rightarrow e(p_2)+\tilde{G}(k_1)+\tilde{G}(k_2)
\end{eqnarray}

In evaluating $|M|^2$ for this process, we note that
the $\gamma\tilde{G}\tilde{G}$ vertex is common to both the direct and the
cross diagrams. Therefore, we evaluate that absolute square first as follows:
\begin{eqnarray}
|V|^2 = \int \frac{d^3k_1}{2k_{10}}\frac{d^3k_2}{2k_{20}}\delta^4(k-k_1-k_2)
\Sigma_s|V_{\gamma\tilde{G}\tilde{G}}|^2
\end{eqnarray}
where 
\begin{eqnarray}
V^{\alpha}_{\gamma\tilde{G}\tilde{G}}= \frac{e M^2}{2\Lambda^4_{\rm S}}
\frac{k_1.k_2}{k^2}\bar{u}(k_1)\gamma^{\alpha}v(k_2)
\end{eqnarray}
Using its gauge invariance, one can write this as
\begin{eqnarray}
|V|^2=A(k^{\alpha}k^{\beta}-g^{\alpha\beta}k^2)
\end{eqnarray}
where $A$ is easily evaluated to be $\frac{2\pi}{3}
\frac{e^4M^4}{4\Lambda^8_{\rm S}}$.
The matrix element $M_{if}$ is given by
\begin{eqnarray}
M_{if}=e^2V^{\alpha}\epsilon^i_{\mu}\bar{u}(p_2)[\gamma^{\alpha}
(\gamma\cdot (q+p_1)-m)^{-1}\gamma^{\mu}+\gamma^{\mu}
(\gamma\cdot (p_2-q)-m)^{-1}\gamma^{\alpha}]u(p_1)
\end{eqnarray}
One can then combine the above equations to write the final formula for 
emissivity as follows:
\begin{eqnarray}
Q_{Comp}=\frac{\alpha^3}{6(2\pi)^7}\left(\frac{M}{\Lambda^2_{\rm S}}\right)^4
\int \frac{d^3p_1d^3p_2d^3q}{p_1p_2q}E_kf(p_1)f_{\gamma}(q)(1-f(p_2))k^2
[\Sigma^3_1X_i/\alpha_i]
\end{eqnarray}
where $\alpha_1=(2q\cdot p_1)^2$, $\alpha_2=(2q\cdot p_2)^2$ and
$\alpha_3=-2q\cdot p_1 q\cdot p_2$
and 
\begin{eqnarray}
X_1=32 (2 m^4 +m^2(2 q\cdot p_1-q\cdot p_2-p_1\cdot p_2)+q\cdot p_1 q\cdot p_2)
\\ \nonumber
X_2=32 (2 m^4 + m^2 (q\cdot p_1-2q\cdot p_2 -p_1\cdot p_2)+ 
q\cdot p_1 q\cdot p_2) \\
\nonumber
X_3= 16 m^2 (q\cdot p_2-q\cdot p_1)+32 p_1\cdot p_2(q\cdot p_1-q\cdot p_2
-p_1\cdot p_2 +2 m^2)
\end{eqnarray}
This integral was evaluated by Monte carlo method to obtain the emissivity as
a function of the supernova core temperature and the parameters $M$ and
$\Lambda_{\rm S}$. Multiplying by the volume of the supernova (with radius
10 kilometers), we obtained the luminosity which was then set less than
$10^{52}$ ergs/sec. The bound on the parameter $\Lambda_{\rm S}$ for
$M=100$ GeV are given table I.

We also wish to point out that we have checked plasma screening effects
on our result by redoing the calculation with the propagator $k^2$
replaced by $k^2+k_{pl}^2$ (with $k_{pl}$ as given in Braaten and Segel
\cite{braaten} in the relativistic limit). We find that these effects are 
well below 5\% level as might be expected from the $k_1\cdot k_2$ factor
in Eq. (9).

\bigskip
\section{Contribution of electron positron annihilation to emissivity}

Using similar methods as in section III, we have calculated the
 emissivity in this case and find
\begin{eqnarray}
Q_{ann}= 8\alpha^2(M/\Lambda^2_{\rm S})^4 T^4 e^{-\mu/T}\mu^5 b(\mu/T)/15\pi^3
\end{eqnarray}
where $b(y)\equiv (5/6)e^yy^{-5}(F^+_5F^-_4+F^+_4F^-_5)$ where 
$F^{\pm}_m(y)=\int dx x^{m-1}/(1+e^{x\pm y})$. Our expression agrees with
that calculated in the Ref.\cite{toldra}. We have adopted their notation
in Eq. (14).
We have evaluated the integrals in the above expression numerically
 and have obtained  lower bounds on the $\Lambda_{\rm S}$
using the requirement that the luminosity in gravitinos have an upper
bound of $10^{52}$ ergs/sec. The results for this case are comparable to
the limits derived from the Compton cooling case and both cases
are collected in Table I. We have varied the core temperature from 30 to
60 MeV and used two typical values for the electron chemical potential 
of 200 and 300 MeV. Again, since the dependence of the luminosity on
the supersymmetry breaking scale goes like the eighth power of 
temperature, for each case
the bigger of the two numbers is the actual bound. We see that the best bound
obtains for the case when the core temperature is assumed to be 60 MeV
as expected and the chemical potential is 200 MeV and is 
\begin{eqnarray}
\Lambda_{\rm S}\geq 500~ GeV
\end{eqnarray}

For the case where $T_c= 50 $ MeV and $\mu=300$ MeV, the bound is 
$\Lambda_{\rm S}\geq 300$ GeV for both the Compton and the annihilation 
cases. Combining these two together we get the bound to be $\Lambda_{\rm S}
\geq 2^{1/8}\times 300=325$ GeV for this choice of supernova parameters. 
Our bound is actually slightly weaker than
the bound given in \cite{toldra} using only the $e^+e^-$ annihilation 
mechanism. Note that we have a higher value for $M$.

Finally, as was first pointed out in \cite{grifols}, the above considerations
 assume that the supernova is transparent to the gravitinos once they are 
emitted. To check this, we have to make sure that the mean free path 
of the grvitino, once emitted is longer than the radius of the supernova
(i.e. 10 km). In Ref.\cite{grifols}, the mean free path was calculated using 
photon-gravitino scattering. Now of course due to the revised Feynman rules,
the whole picture is different. The main contribution to opacity comes
from gravitino scattering off protons and electrons. The mean free path
for $\Lambda_{\rm S}= 300$ GeV due to proton scattering has been calculated
in \cite{toldra} and we do not repeat their calculation. They find that
for $\Lambda_{\rm S}\leq 220$ GeV, the gravitinos get trapped.
Using the techniques employed in \cite{barb}, they have shown that the
emissivity remains too large until $\Lambda_{\rm S}\leq 70$ GeV, so that
the excluded range of $\Lambda_{\rm S}$ is between 70 GeV and 300- 500 GeV
depending on the choice of core temperature and the electron chemical 
potential. This can be translated to a forbidden range for the gravitino mass
of $ 10^{-7}~eV \leq m_{\tilde{G}}\leq .6-6 \times 10^{-6}$ eV. This 
bound is very similar to the bound obtained in \cite{grifols}.

\section{Neutron star cooling constraints }
	Neutron star cooling is an attractive area in which to test
non-standard particle physics models, in principle, because the standard model
of neutron cooling, based on the "modified URCA process," gives insufficient
cooling to match current observations.  For a recent review and refernces to
the literature see \cite{page}.  Briefly, for neutron stars
of ages between about $10^2$ and $10^5$ years, the standard model (slow
cooling) is one of neutrino emission from an isothermal superfluid core with an
energy gap in the 100 keV range.  However data from two pulsars of ages around
$10^4$ years show temperatures a factor of 5 or so below the standard model
prediction.  Time is measured by $P/(2\frac{dP}{dt})$ where $P$ is the pulsar
period while temperature is measured from the (X-ray) black body spectrum
(keeping in mind that the surface temperature is believed to be roughly the 
square root of the interior temperature (in eV)).  For earlier times only a few
upper bounds are available so that there is no good evidence on the temperature
dependence of the mechanism that is providing the extra (fast) cooling. 
 However,
considerable extra cooling is required since the URCA cooling rate is
proportional to the eighth power of the temperature.

	Various mechanisms to provide the extra cooling have been proposed. 
These include a smaller superfluid energy gap, meson condensates, and nucleon
dissociation into quarks under the high central pressure.  Were gravitino
emission the source of the extra cooling, that fact would provide a measurment
of gravitino coupling rather than a bound.  Thus we turn to the question of
whether that possibility can be entertained.

As noted earlier, the effective coupling of the gravitino pair to electrons
and quarks is via the photon exchange -- however for superlight gravitinos,
it reduces effectively to a four Fermi interaction like the weak Fermi
coupling except that the coupling only occurs with electrically charged
particles. One channel for neutron star cooling that involves gravitino
emission is therefore via a neutral current like coupling with the
coupling constant given by $q_ae^2M^2/(4\Lambda^4_{\rm S})$. Thus the
relevant process is the analog of the neutral current URCA process
$n+p\rightarrow n+p+\tilde{G}\tilde{G}$. The temperature dependence of this 
process will be $T^8$ as in the URCA process. So it can compete
with the neutrino cooling of neutron stars in the appropriate range
for the coupling parameters. Assuming that neutrino luminosity describes
the cooling of neutron stars long after their birth\cite{page}, we
assume that luminosity via gravitinos is of same order of magnitude.
(We choose this arbitrarily in view of the scant data
on neutron stars at the moment.)

As is well 
known, in the case of the neutron stars the neutral current driven URCA
process is about a factor of 30 lower\cite{friman,raffelt}
than the corresponding charged current process, of which a factor
of four comes from the coupling and the rest from phase space. Since our
couplings are given, we demand that gravitino cooling be
less than the full neutrino URCA process luminosity.
This leads us to the bound that
\begin{eqnarray}
\frac{4\pi\alpha M^2}{2\Lambda^4_{\rm S}}\leq G_F/\sqrt{2}
\end{eqnarray}
This implies that $\Lambda_{\rm S}\geq 200$ GeV, which is comparable to the
 bounds obtained from the supernova. Considerably larger coupling,
contrary to the supernova bounds derived above, would be required for
gravitino emission to provide an explanation for the fast neutron star
cooling observed. 
                                                                   
\section{Conclusion}
In conclusion, we have revisited the issue of supernova and neutron star 
constraints on the supersymmetry breaking scale and the ensuing constraint
on the mass of the superlight gravitino using the revised Feynman rules for
the goldstino coupling to matter in a large class of supersymmetric models
with a low SUSY breaking scale.
We find that the lower limits on the supersymmetry breaking scale 
$\Lambda_{\rm S}$ are in the range of 200 to 500 GeV and the resulting
lower limit on the gravitino mass is $.6-6\times 10^{-6}$ eV. One must 
however remember that while the form of the gravitino coupling to matter
is universal,
there is a model dependent parameter $M$ that characterises
the strength of the coupling.
In any case, these bounds are comparable to the collider bounds
obtained by various authors\cite{roy2}.
 As has been noted earlier\cite{riotto},
the corresponding constraints on the gravitino mass are much more severe
in models where the gravitino is also accompanied by superlight scalar/or
pseudoscalar particles.
\bigskip

\noindent{\Large \bf{Acknowledgements}}

\bigskip

The work of D. A. D. is supported in part by the U. S. Department of Energy
under Grant No. DE-FG013-93ER40757;
the work of R. N. M.
is supported by the National Science Foundation under Grant No. PHY-9421386.
V. L. T. would like to thank
David Branch and Eddie Baron for communications regarding supernovae.
 R. N. M. would like to thank Markus Luty for discussions.

\bigskip

\begin{center}
{\bf Table I}
\end{center}

\begin{tabular}{|c||c||c||c|}\hline
$T_c$  & $\mu_e$  & $\Lambda^{min}_{\rm S}$ 
 from $Q_{ann}$ & $\Lambda^{min}_{\rm S}$ from $Q_{Comp}$ \\ \hline
30 & 200 & 165 & 175 \\ \hline
30 & 300 & 135 & 160 \\ \hline
40 & 200 & 250 & 230 \\ \hline
40 & 300 & 225 & 240 \\ \hline
50 & 200 & 335 & 300 \\ \hline
50 & 300 & 305 & 300 \\ \hline
60 & 200 & 420 & 500 \\ \hline
60 & 300 & 400 & 385 \\ \hline
\end{tabular}

\noindent{\bf Table caption}: The lower bounds on the supersymmetry breaking
scale $\Lambda_{S}$ in GeV derived from the two dominant processes in the
supernova for various values of core temperature $T_c$ in MeV and electron
chemical potential $\mu_e$ in MeV. We have chosen $M=100$ GeV.

\bigskip

\noindent{\bf Figure caption} Feynman diagram for  Compton cooling where
the lower solid line denotes the electron, the wavy lines denote the photon
and upper external solid lines denote the gravitino pair.
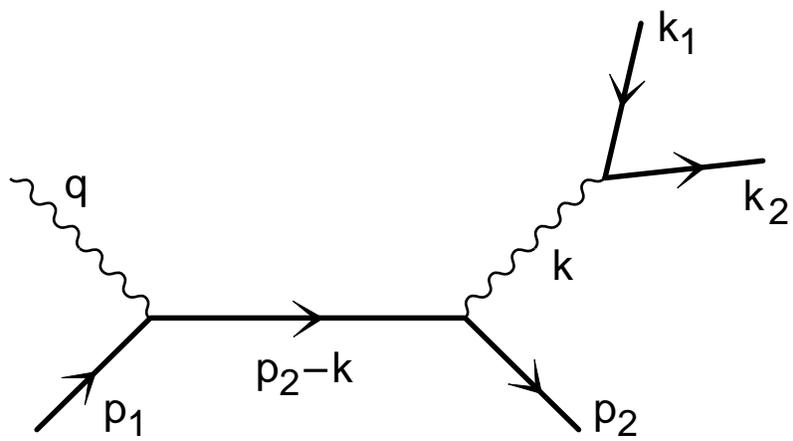
\end{document}